\documentstyle[12pt]{article}
\newlength{\extraspace}
\newlength{\extraspaces}
\catcode`\@=11
\def\numberbysection{\@addtoreset{equation}{section}
\def\theequation{\arabic{section}.\arabic{equation}}}

\begin{document}
\addtolength{\baselineskip}{.7mm}
\thispagestyle{empty}
\begin{flushright}
TIT/HEP--340 \\
UCLA/96/TEP/39  \\
%{\tt hep-th/9603206} \\
July, 1996 
\end{flushright}
\vspace{2mm}
%
%\hfill{UCLA/96/TEP/39}
%\vspace{24pt}
%
\begin{center}
{\large{\bf   Gauge Symmetry Breaking through Soft Masses}}
\vspace{5pt}
{\large{\bf   in Supersymmetric Gauge Theories }} \\[15mm]
{\sc Eric D'Hoker}\footnote{
\tt e-mail: dhoker@physics.ucla.edu}   \\[2mm]
{\it Department of Physics and Astronomy, \\
 University of California at Los Angeles\\
Los Angeles, CA 90024, USA} \\[4mm]
{\sc Yukihiro Mimura}\footnote{
\tt e-mail: mim@th.phys.titech.ac.jp}  \hspace{2mm} 
 and  \hspace{2mm}
{\sc Norisuke Sakai}\footnote{
Speaker 
\tt e-mail: nsakai@th.phys.titech.ac.jp}   \\[2mm]
{\it Department of Physics, Tokyo Institute of Technology \\
Oh-okayama, Meguro, Tokyo 152, Japan}  \\[10mm]
{\bf Abstract}\\[5mm]
{\parbox{14cm}{\hspace{5mm}
Effects of soft breaking terms in $N=1$ supersymmetric gauge 
theories are studied. 
{}For $N_f<N_c$, we include the dynamics of the 
non-perturbative superpotential and use the original (s)quark and 
gauge fields. 
{}For $N_f>N_c+1$, we formulate the dynamics in 
terms of dual (s)quarks and a dual gauge group $SU(N_f-N_c)$. 
The mass squared of squarks can be negative triggering the 
 spontaneous breakdown of flavor and color symmetry. 
The general condition for the stability of the vacuum is derived. 
We determine this breaking pattern, derive the spectrum, and argue that 
the masses vary smoothly as one crosses from the Higgs phase into the 
confining phase exhibiting the complmentarity. 
 }}
\end{center}
\vfill
\newpage
\setcounter{section}{0}
\setcounter{equation}{0}
\setcounter{footnote}{0}
\def\theequation{\arabic{section}.\arabic{equation}}
%
%
%%%%%%%  Section 1  %%%%%%%%%%%%%%%%%%%%%%%%%%%%%%%%%%%%%%%%
%
\vspace{7mm}
\pagebreak[3]
\addtocounter{section}{1}
\setcounter{equation}{0}
\setcounter{subsection}{0}
\setcounter{footnote}{0}
\begin{center}
{\large {\bf \thesection. Introduction}}
\end{center}
\nopagebreak
\medskip
\nopagebreak
\hspace{3mm}
Supersymmetric theories has now become a standard theory to solve the 
gauge hierarchy problem and to unify all the forces in nature. 
Recent advances to understand the dynamics of supersymmetric 
gauge theories has provided a clear picture of nonperturbative 
effects  \cite{VeYa} -- \cite{SeWi}. 

It has been anticipated that these features are not restricted to 
$N=1 $ supersymmetric gauge theories, but should survive -- 
at least in a qualitative way -- to non-supersymmetric gauge theories. 
The models in which this proposal is perhaps most easily verified 
are those where supersymmetry is spontaneously 
broken (through the introduction of an additional sector of 
fields \cite{FaIl} -- \cite{Oraifeartaigh}) or 
those where soft, explicit supersymmetry breaking terms are added as a 
perturbation on the gauge dynamics \cite{GiGr}. 
The latter scheme of supersymmetry breaking was used in the original 
proposal of supersymmetric grand unified theories \cite{Sakai}, 
\cite{DiGe} and provides a general framework for 
the usual formulation of the Minimal Supersymmetric 
Standard Model \cite{Nilles}. 

In a series of papers, it was argued that the addition of 
perturbative, soft 
supersymmetry breaking mass terms, with $m^2 \geq 0$, essentially 
preserves the 
qualitative picture of the dynamics derived for $N=1$ supersymmetric 
QCD (SQCD) \cite{AhSoPe}. 
An effective low energy theory is used 
in terms of color singlet meson and (for $N_f\geq N_c$) 
baryon fields, appropriate for the confining phase, and 
the effects of the non-perturbative 
superpotential of Affleck, Dine and Seiberg \cite{AfDiSe} are included. 

More recently we have we investigate $N=1$ supersymmetric QCD, 
again with soft supersymmetry breaking mass terms added, but this time 
with $m^2<0$  for at least some of the squark fields. 
In the present paper, we will briefly report our findings. 
{}For general values of the soft 
supersymmetry breaking mass terms with $m^2<0$, the Hamiltonian will 
become unbounded from below, thus destabilizing the entire theory.  
{}For certain simple ranges of the masses, however, we show that 
stable vacua exist through a balance between the soft supersymmetry 
breaking mass terms, the quartic $D^2$ term 
for the squark fields and ( for $N_f < N_c$ ) the 
non-perturbative effective potential. 
We argue that in these solutions, flavor as well as color symmetry 
are spontaneously 
broken through the vacuum expectation value of the squarks. 
We consider the most general soft supersymmetry breaking terms 
respecting $R$-symmetry, for simplicity. 

{}For the analysis of spontaneous flavor and color symmetry breakdown 
when $N_f<N_c$, we formulate the dynamics in terms of the fundamental
quark superfields instead of in terms of meson superfields, in 
contrast with the analysis in ref.\cite{AhSoPe}. 
This choice appears more natural when dealing with the theory in the
Higgs phase, 
rather than in the confining phase. 
In fact, the original calculation of the nonperturbative superpotential 
was justified precisely by considering gauge symmetry breakdown due to 
the vacuum expectation values of squark fields \cite{AfDiSe}. 

{}For small values of the soft supersymmetry breaking mass terms, we 
expect the confining and Higgs phases to be smoothly connected 
to one another, with matching low energy spectra. 
We find that this principle of 
complementarity between the confining and Higgs phases \cite{FrSh}
can indeed be satisfied in these theories. To do so however, 
it is necessary to consider the nonminimal kinetic terms for color 
singlet fields in the confining phase. 
This is in contrast to the minimal kinetic term advocated in 
ref.\cite{AhSoPe}. 

{}For large values of the soft supersymmetry breaking mass terms, we
expect the semi-classical spectrum for the Higgs phase, 
derived in terms of the fundamental quark superfields, to remain reliable.
Thus, we shall calculate the 
semi-classical spectrum for this model for all ranges of soft 
breaking mass terms. 
{}For $N_f > N_c +1 $, we will use the dual variables of ref.\cite{Seiberg}, 
which are more appropriate to describe the spontaneous symmetry 
breakdown of flavor and (dual) color symmetry. 

{}Finally, we note that there is a simple extension of standard SQCD, 
obtained by gauging also an additional anomaly-free $U(1)_X$ symmetry, 
and adding a Fayet-Illiopoulos $D$-term 
\cite{FaIl} for the corresponding gauge multiplet. 
(The simplest case would be where this $U(1)_X$ is just baryon number 
symmetry, but any anomaly-free 
$U(1)_X$ would do.) 
No explicit supersymmetry breaking terms are added; instead, a 
supersymmetric mass term is included in the superpotential, which 
stabilizes the vacuum. 
In this model, supersymmetry is broken spontaneously, and 
soft mass terms with $m^2<0$ automatically arise from 
the Fayet-Illiopoulos $D$-term.  
This model provides an economical realization of some of the
effects of soft supersymmetry breaking mass terms generated directly by 
spontaneous breakdown of  supersymmetry. 
It will be discussed in a companion paper.
A different model has been analyzed which obtains the soft breaking 
terms from the spontaneous breakdown of supersymmetry \cite{EvHsSc}. 

%
%%%%%%%  Section 2  %%%%%%%%%%%%%%%%%%%%%%%%%%%%%%%%%%%%%%%%
%
\vspace{7mm}
\pagebreak[3]
\addtocounter{section}{1}
\setcounter{equation}{0}
\setcounter{subsection}{0}
\setcounter{footnote}{0}
\begin{center}
{\large {\bf \thesection. Dynamics for $N_f < N_c$
}}
\end{center}
\nopagebreak
\medskip
\nopagebreak
\hspace{3mm}
In this Section, we shall consider supersymmetric Yang-Mills theory 
with gauge group $SU(N_c)$ and $N_f$ flavors of squarks and quarks 
(with $N_f < N_c$), transforming under the representation 
$N_c\oplus \bar  N_c$ of $SU(N_c)$. This 
theory is the natural supersymmetric extension of QCD, 
and will be referred to as SQCD. 
The corresponding chiral superfields 
\begin{equation}
\hat Q_a{}^i \; 
\qquad 
\hat {\bar Q} _i{}^a
\qquad \qquad
a=1, \cdots, N_c;
\qquad
i=1, \cdots , N_f\; ,
\label{eq:one}
\end{equation} 
contain the squark fields $Q$ and $\bar Q$ and the left-handed quark 
fields
$\psi _Q$ and $\psi _{\bar Q}$ respectively. There is a natural color 
singlet
meson chiral superfield $\hat T$, defined by
\begin{equation}
\hat T_i{}^j = \hat {\bar Q}_i{}^a \hat Q_a{}^j
\label{eq:two}
\end{equation}
with scalar components $T_i{}^j$. 
Superfields are denoted by a cap on the scalar components. 

As a starting point we consider classical massless SQCD 
whose Lagrangian ${\cal L}_0$ is 
determined by $SU(N_c)$ gauge invariance, 
by requiring that the 
superpotential for the quark superfields vanish identically :
\begin{equation}
{\cal L}_0 = \int d^4 \theta {\rm \, tr} \{
\hat Q ^{\dagger}  e^{2g\hat V}\hat Q + 
\hat {\bar Q}  e^{-2g\hat V}\hat{\bar Q} ^{\dagger} \} 
+ \frac{1}{2} \int  d^2\theta ~{\rm \, tr} W W 
+ \frac{1}{2} \int  d^2\bar \theta ~{\rm \, tr} \bar W \bar  W 
\label{eq:three}
\end{equation}
This theory has a global symmetry, 
$G_f=SU(N_f)_Q\times SU(N_f)_{\bar Q} \times
U(1)_B\times U(1)_R$, 
with $R$-charges ( baryon number) are 
given by $1-N_c/N_f$ ($1$) for $Q$, and 
$1-N_c/N_f$ ($-1$) for $\bar Q$. 
\par
Exact nonperturbative results in supersymmetric gauge theories 
can be given for the $F$-type term which is a chiral superspace 
integral of a superpotential $W_{NP}$ \cite{AfDiSe} of 
the quark superfields $\hat Q$ and $\hat {\bar Q}$ given as follows
\begin{equation}
\int d^2 \theta W_{NP}(\hat Q, \hat {\bar Q}) = (N_c-N_f) \Lambda
^{3+2N_{f}/(N_c-N_f)}
\int d^2\theta ~({\rm det} \hat {\bar Q}\hat Q ) ^{-1/(N_c - N_f)} 
\label{eq:seven}
\end{equation}
%
%%%%%%%%%%%%%%%%%%%%%sect.2.1%%%%%%%%%%%%%%%%%%%%%%
%
%
We choose to break supersymmetry explicitly, by adding to the 
Lagrangian ${\cal L}_0$ soft supersymmetry breaking terms 
 \cite{GiGr} for the quark supermultiplet. 

{}For the sake of simplicity, we shall add to the  Lagrangian only soft 
supersymmetry breaking squark mass terms and  neglect effects due to 
gaugino masses and supersymmetric flavor masses.   
Generic mass squared for squark and antisquarks are given by matrices 
$M^2_Q$ and $M^2_{\bar Q}$ 
\begin{equation}
{\cal L} _{sb} = - 
\{ {\rm \, tr} Q M_{Q}^{2} Q^\dagger  + {\rm \, tr} 
\bar Q^\dagger M^2_{\bar Q} \bar Q 
%m_{\bar Q_{i}} ^{2} \bar Q^\dagger {} _{a} {}^{i} \bar Q_{i} {}^{a}
 \}
\label{eq:six}
\end{equation}
As we remarked above, when $M^2_Q$ and $M^2_{\bar Q}$ 
are proportional to the identity matrix, 
the global flavor symmetry is unchanged~: $SU(N_f)_Q \times
SU(N_f)_{\bar Q}\times U(1)_B
\times U(1)_R$.

When either $M^2_Q$ or $M^2_{\bar Q}$ is not positive definite, 
we expect the pattern of symmetry breaking to be
substantially  different. Global flavor symmetry should be 
spontaneously broken,
and $Q$ and/or $\bar Q$ should acquire non-vanishing vacuum 
expectation values.
These non-zero vacuum expectation values, in turn, are expected 
to break color
$SU(N_c)$ and give mass to some of the gauge particles through 
the Higgs
mechanism. This is the so-called Higgs phase.  

According to standard lore, (originally derived from lattice gauge 
theory) the confining and Higgs phases are smoothly connected to one 
another in at least some region of parameter space \cite{FrSh}. 
There should be a one to one correspondence between the 
observables in both phases, suggesting that -- in principle -- 
color singlet meson fields could still be used to describe the
dynamics of the Higgs phase.  
In practice, however, a formulation in terms of
colored fields appears more suitable instead. 
Indeed, physical free quarks and certain massive gauge bosons 
are expected to appear in the low energy spectrum, and  
it is unclear how to represent these degrees of freedom in 
terms of meson variables. 
Thus, we shall use the original squark $Q,~\bar Q$, 
quark $\psi _Q, ~ \psi _{\bar Q}$, and gauge boson and fermion fields 
as physical variables at low energy. 
%
%
%%%%%%%%%%%%%%%%%%%%%sect.2.2%%%%%%%%%%%%%%%%%%%%%%
%
Without SUSY breaking soft masses, the vacuum is known to runaway. 
{}For generic matrices $M^2_Q$ and $M^2_{\bar Q}$, 
we find the stability condition for the vacuum as 
\begin{equation}
m_{Q_i}^2 + m_{\bar Q_j}^2 \geq 0 %\qquad \qquad i=1,\cdots N_f
\label{eq:nine}
\end{equation}
for any pair of $i, j=1,\cdots N_f$. 

%%%%%%%%%%%%%%%%%%%%%sect.2.3%%%%%%%%%%%%%%%%%%%%%%
%

{}For simplicity, we explicitly analyze only the case where the 
mass squared for all $Q$'s and $\bar Q$'s are equal to $-m_{Q}^2$ 
and $m_{\bar Q}^2$ respectively, thus preserving the entire global 
symmetry 
$SU(N_f)_Q\times SU(N_f)_{\bar Q} \times U(1)_B \times U(1)_R$. 
After a somewhat lengthy analysis, we find that there is a unique 
minimum of the potential at which the first $N_f\times N_f$ block 
of $Q$ and $\bar Q$ are nonvanishing and are proportional to the 
identity 
\begin{equation}
\langle 0 | Q | 0 \rangle = 
\left ( \begin{array}{c} Q_{0} I_{N_f} \\ 0 \\ \end{array} \right ) 
\qquad \qquad \qquad 
\langle 0 | \bar Q | 0 \rangle = 
\left ( \begin{array}{cc} \bar Q_{0}I_{N_f} & 0 \\ \end{array} \right )
\label{eq:twentyfive}
\end{equation}
. 
The minimum conditions are
\begin{eqnarray}
0 &=&
(\gamma -1) Q_{0} ^{-2\gamma}   \bar Q_{0}^{-2\gamma}
  +\gamma     Q_{0} ^{-2-2\gamma} \bar Q_{0}^{2-2\gamma} 
  -\frac{g^2}{2\gamma} (Q_{0}^{2}-\bar Q_{0}^{2}) +m_{Q}^{2} 
\nonumber \\
0 &=&
(\gamma -1) Q_{0} ^{-2\gamma}   \bar Q_{0}^{-2\gamma}
  +\gamma     Q_{0} ^{2-2\gamma} \bar Q_{0}^{-2-2\gamma} 
  +\frac{g^2}{2\gamma} (Q_{0}^{2}-\bar Q_{0}^{2}) -m_{\bar Q}^{2} 
\label{eq:twentyone}
\end{eqnarray}

%
%%%%%%%%%%%%%%%%%%%%%sect.2.4%%%%%%%%%%%%%%%%%%%%%%
%
%\subsection{Spectrum and Unbroken Symmetries}
%
We have obtained the spectra in this vacuum. 
As anticipated, the scalar particles contain massless Nambu-Goldstone 
boson corresponding to the spontaneous breakdown of $U(1)_R$ 
and $SU(N_f)$ global symmetry.

%%%%%%%%%%%%%%%%%%%%%sect.2.5%%%%%%%%%%%%%%%%%%%%%%
\vspace{7mm}
\pagebreak[3]
\addtocounter{section}{1}
\setcounter{equation}{0}
\setcounter{subsection}{0}
\setcounter{footnote}{0}
\begin{center}
{\large {\bf \thesection. 
Complementarity between confining and Higgs phase }}
\end{center}
\nopagebreak
\medskip
\nopagebreak
\hspace{3mm}
According to the complemetarity argument, one should be able to relate 
our mass spectra calculated in terms of colored elementary fields 
with the result in ref.~\cite{AhSoPe} where the meson fields 
$T$ are used as fundamental variables in the low energy effective theory. 
They have assumed a standard minimal kinetic term for the meson fields, 
because the K\"ahler potential can not be constrained by holomorphy. 
Consequently their results on mass spectra differed from ours, 
although overall qualitative picture was the same. 

On the other hand, if we take large VEV for squark fields, 
we should be in a nearly perturbatiove region. 
Therefore there should be the correct K"ahler potential for the 
color singlet meson which correspond to our choice of perturbative 
kinetic term for squarks. 
We indeed find that a similar analysis as in ref.~\cite{AhSoPe} 
with the K\"ahler potential for meson 
\begin{equation}
K[T]=  2{\rm \, tr} \left( T^\dagger T \right)^{1 \over 2}
\end{equation}
reproduces our mass spectra correctly. 

This shows that the complementarity is also valid in supersymmetric 
situation, and the higher order terms in K\"ahler potential are 
important when some of the fields acquire VEV. 

%
%%%%%  Section 3  %%%%%%%%%%%%%%%%%%%%%%%%%%%%%%%%%%%%%%%%
\vspace{7mm}
\pagebreak[3]
\addtocounter{section}{1}
\setcounter{equation}{0}
\setcounter{subsection}{0}
\setcounter{footnote}{0}
\begin{center}
{\large {\bf \thesection. Dynamics for $N_f > N_c+1$ }}
\end{center}
\nopagebreak
\medskip
\nopagebreak
\hspace{3mm}
%
%\newsection{ Dynamics for $N_f > N_c$}

We have aslo analyzed the case of $N_f > N_c+1$ using the 
dual description. 
Qualitaive picure is similar to the previous case of 
$N_f < N_c+1$. 

The dual description for the gauge group $SU(N_c)$
and $N_f$ flavors of quarks and antiquarks (with $N_f > N_c+1 $) 
has a gauge group $SU(\tilde{N}_c)$ with $N_f$ flavors,
where $\tilde{N}_c = N_f - N_c$. The elementary chiral superfields
in the dual theory are dual quark $\hat q$ 
and meson $\hat T$ superfields,
$
\hat{q}^a{}_i \ \  \hat{\bar{q}}^i{}_a \ \ \hat{T}^i{}_j \\ 
a=1, \cdots, \tilde{N}_c; \ \ i,j=1, \cdots N_f
$
.

Since $\hat q$, $\hat {\bar q}$ and $\hat T$ are effective fields, 
their kinetic terms need not have canonical normalizations; 
in particular, they can receive nonperturbative quantum corrections. 
Thus, we introduce into the 
(gauged) K\"{a}hler potential for $\hat q$, $\hat {\bar q}$ and $\hat T$
normalization parameters
$k_q$ and $k_T$ as follows  
\begin{equation}
K[\hat q,\hat {\bar q},\hat T,\hat v]
= k_q {\rm tr} (\hat q^{\dagger} e^{2\tilde
g \hat v} \hat q  + \hat {\bar q} e^{-2\tilde g \hat v} {\bar
q}^{\dagger})
                 + k_T {\rm tr} \hat T^{\dagger} \hat T.
\label{kaehler}
\end{equation}
Here, we denote by $\hat v$ the $SU(\tilde N_c)$ color gauge superfield, and by
$\tilde g$ the associated coupling constant.  (Pure gauge terms will not be
exhibited explicitly.) In principle, these normalization parameters
are determined by the dynamics of the underlying microscopic theory.
{}Furthermore, it has been pointed out in \cite{Seiberg} that a superpotential
coupling $q$, $\bar q$ and $T$ should be added as follows 
\begin{equation}
W= \hat{q}^a{}_i \hat{T}^i{}_j \hat{\bar{q}}^j{}_a.
\end{equation}

We add soft supersymmetry breaking terms 
to the Lagrangian for the dual quark and meson supermultiplets.
{}For simplicity we shall assume that $R$-symmetry is maintained so 
that neither $A$-terms  nor gaugino masses are present in the 
Lagrangian. 

%
%\subsection{Vacuum Stability}

When the eigenvalues of $M_q^2$, $M_{\bar q}^2$ and $M_T^2$ can take 
generic positive or negative values, the scalar potential may be 
unbounded from below.
A necessary condition for which the potential is bounded from below is 
that $M_T^2$ be a positive definite matrix.
This is because there is no quartic term of $T$.

The D terms vanish when the vacuum expectation values are given by:
\begin{equation}
\langle 0| q |0\rangle = \left(
 \begin{array}{ccccc}
 q_1 &        &                 &   & \\
     & \ddots &                 & 0 & \\
     &        & q_{\tilde{N}_c} &   &
 \end{array}
\right), \qquad
\langle 0| {\bar q}|0 \rangle = \left(
 \begin{array}{ccc}
 \bar{q}_1 &        &                       \\
           & \ddots &                       \\
           &        & \bar{q}_{\tilde{N}_c} \\
           &        &                       \\
           &   0    &   
 \end{array}
\right), 
\end{equation}
with the combinations $|q_i|^2 - |\bar{q}_i|^2$ independent of $i$.

If we set the squark masses to be zero,
the space where $|q_i|^2$ is independent of $i$ and 
$\bar{q}=0$ is a subspace of the moduli space of vacua. 
If we insist on flavor symmetric mass squared matrix and on having a 
negative eigenvalue, we are forced to have a potential unbounded 
from below. In fact, in the next subsection, we shall establish more generally
that to have a potential bounded from below, we must have 
\begin{equation}
m_1^2 + \cdots + m_{\tilde{N}_c}^2 \geq 0,
\end{equation} 
where $m_i^2$ are eigenvalues of the matrix $M_q^2$ or $M_{\bar q}^2$,
and they are set to be $m_1^2 \leq m_2^2 \leq \cdots \leq m_{N_f}^2$.

Therefore we consider the simplest stable situation, where 
the $n$ eigenvalues of $M^2_q$ is negative and same, 
while all the others are positive or
zero. The $n$ should be smaller than $\tilde{N}_c$.
{}For simplicity we shall also assume that the soft supersymmetry 
breaking positive mass squared terms for squarks have a flavor symmetry 
$SU(N_{f}-n)_Q \times SU(N_{f})_{\bar Q}$. 
As a result, the $N_{f}-n$
positive  eigenvalue of $M_q^2$ are all the same, while the $N_f$ 
eigenvalues of  $M_{\bar q}^2$ are the same 
:$M_{\bar q}^2=m_{\bar q}^2 I_{N_{f}}$. 
We also assume 
%and 
$M_T^2{}^i{}_j{}^k{}_l=m_T^2 \delta^i_j \delta^k_l$.

After all, we find that there are only two solutions for possible 
minimum,
described as follows. 
\begin{enumerate}
\item Only  $q_1, \cdots, q_n \neq 0$, while $q_i =0, ~i > n$ 
and $\bar q_i=0$ for all $i$.
The values of $q_1, \cdots, q_n$ are the same. We call the common value 
as $q_0$.
The value of $q_0$ and the potential in this configuration are
given by
\begin{equation}
q_0^2 = \frac2{\tilde g^2 \tilde{\gamma}} m_{q_1}^2,
\qquad \qquad
V= - \frac{n}{\tilde g^2 \tilde{\gamma}} m_{q_1}^4.
\end{equation}
where
\begin{equation}
\tilde{\gamma} = \frac{\tilde{N}_c-n}{\tilde{N}_c}.
\end{equation}

\item Only $q_1, \cdots, q_n \neq 0$ and $\bar{q}_1, \cdots \bar{q}_n \neq 0$,
while  $q_i=\bar q_i=0, ~i > n$. 
The values of $q_0 \equiv q_1=\cdots=q_n$ and $\bar q_0\equiv \bar{q}_1=
\cdots \bar{q}_n$ are then given by
\begin{equation}
\left( \begin{array}{c} q_0^2 \\  \\ \bar{q}_0^2 \end{array} \right)
= \frac{1}{\tilde{\gamma} \tilde g^2 - \frac1{k_T}}
\left( \begin{array}{c}
\frac12 \tilde{\gamma} \tilde g^2k_T (m_{q_1}^2- m_{\bar q}^2)
        +  m_{\bar q}^2 \\ \\
\frac12 \tilde{\gamma} \tilde g^2k_T (m_{q_1}^2- m_{\bar q}^2)
        -  m_{q_1}^2 \end{array}
\right)
\label{vevofq}
\end{equation}
Given the fact that $q_1^2$ and $\bar q_1^2$ must be positive, this
expression yields a solution only when the following condition is 
satisfied
\begin{equation}
%m_{q_1}^2- m_{\bar q}^2 \geq 0
%\qquad\qquad
\frac12 \tilde{\gamma} \tilde g^2 k_T (m_{q_1}^2- m_{\bar q}^2) \geq  m_{q_1}^2.
\label{eq:threethirty} 
\end{equation}
The value of the potential at the stationary point is given by 
\begin{equation}
V=- \frac{n}4 \frac{\tilde{\gamma} k_T \tilde g^2}
         {\tilde{\gamma} \tilde g^2 - \frac1{k_T}}
    \left( m_{\bar q}^2 - \frac{\frac12 \tilde{\gamma} \tilde g^2 
               - \frac1{k_T}}{\frac12 \tilde{\gamma} \tilde g^2 }
      m_{q_1}^2 \right)^2 -\frac{n}{ \tilde{\gamma}\tilde g^2}
      m_{q_1}^4
\end{equation}

\end{enumerate}

Therefore, whenever conditions (\ref{eq:threethirty}) is satisfied, 
solution 2 is the absolute minimum of the potential and describes the 
true ground state.  If condition (\ref{eq:threethirty}) is not 
satisfied, solution 1 is the absolute minimum.

In our model, the Lagrangian has a global 
$SU(N_f-n)_Q \times SU(n)_Q \times U(1)_Q
\times SU(N_f)_{\bar Q} \times U(1)_B \times U(1)_R$ 
symmetry.
%When $m_{q_1}^2 <m_{\bar q}^2$, or 
When the coupling $\tilde g$ is too weak, the
condition (\ref{eq:threethirty}) is not satisfied, solution 1. is 
the absolute minimum, and flavor symmetry is not broken. 

On the other hand, when the gauge coupling is 
strong, the condition (\ref{eq:threethirty}) is satisfied. 
Therefore solution 2 is the absolute minimum, and flavor symmetry is 
spontaneously broken as follows;
\begin{eqnarray}
&&SU(N_f-n)_Q \times SU(n)_Q \times U(1)_Q
\times SU(N_f)_{\bar Q} \times U(1)_B \times U(1)_R 
\nonumber \\
&& \qquad 
\longrightarrow SU(N_f-n)_Q \times SU(N_f-n)_{\bar Q}
\times SU(n)_V \times U(1)_V 
\times U(1)_{B'} \times  U(1)_{R'},
\end{eqnarray}
where $SU(n)_V$ is the diagonal subgroup of $SU(n) \subset
SU(\tilde{N}_c)$ and $SU(n)_Q \times SU(n)_{\bar Q}$.
The spontaneous breaking of the global symmetry induces spontaneous 
breaking of color gauge symmetry  
$SU(\tilde{N}_c) \to SU(\tilde{N}_c-n)$.
One interesting feature of the present case is that the gauge symmetry 
is broken without chiral symmetry breaking. 
We find Nambu-Goldstone bosons for spontaneous breakdown of 
$SU(N_f)/SU(N_f-n)$. 
%
%%%%%%%  References  %%%%%%%%%%%%%%%%%%%%%%%%%%%%%%%%%%%%%%%
\vspace{5mm}
%\newpage
%
\newcommand{\NP}[1]{{\it Nucl.\ Phys.\ }{\bf #1}}
\newcommand{\PL}[1]{{\it Phys.\ Lett.\ }{\bf #1}}
\newcommand{\CMP}[1]{{\it Commun.\ Math.\ Phys.\ }{\bf #1}}
\newcommand{\MPL}[1]{{\it Mod.\ Phys.\ Lett.\ }{\bf #1}}
\newcommand{\IJMP}[1]{{\it Int.\ J. Mod.\ Phys.\ }{\bf #1}}
\newcommand{\PRP}[1]{{\it Phys.\ Rep.\ }{\bf #1}}
\newcommand{\PR}[1]{{\it Phys.\ Rev.\ }{\bf #1}}
\newcommand{\PRL}[1]{{\it Phys.\ Rev.\ Lett.\ }{\bf #1}}
\newcommand{\PTP}[1]{{\it Prog.\ Theor.\ Phys.\ }{\bf #1}}
\newcommand{\PTPS}[1]{{\it Prog.\ Theor.\ Phys.\ Suppl.\ }{\bf #1}}
\newcommand{\AP}[1]{{\it Ann.\ Phys.\ }{\bf #1}}
\newcommand{\ZP}[1]{{\it Zeit.\ f.\ Phys.\ }{\bf #1}}
\end{document}